\pdfoutput=1

\documentclass[11pt,twoside,a4paper,cmspaper,final,collab]{cms-tdr}
\def\svnVersion{a5adf79-D}\def\svnDate{2019/12/17}\def\cmsCernNoTag{CERN-EP-2019-183}\def\cmsCernDate{\today}\def\cmsMessage{"Published in Physics Letters B as \href{http://dx.doi.org/10.1016/j.physletb.2020.135285}{\doi{10.1016/j.physletb.2020.135285}.}"}

\begin{document}\cmsNoteHeader{TOP-18-011}

\hyphenation{had-ron-i-za-tion}
\hyphenation{cal-or-i-me-ter}
\hyphenation{de-vices}

\newlength\cmsFigWidth
\ifthenelse{\boolean{cms@external}}{\setlength\cmsFigWidth{0.49\textwidth}}{\setlength\cmsFigWidth{0.65\textwidth}}
\ifthenelse{\boolean{cms@external}}{\newcommand{\cmsLeft}{upper\xspace}}{\newcommand{\cmsLeft}{left\xspace}}
\ifthenelse{\boolean{cms@external}}{\newcommand{\cmsRight}{lower\xspace}}{\newcommand{\cmsRight}{right\xspace}}

\cmsNoteHeader{TOP-18-011}
\title{Measurement of the $\ttbar\bbbar$ production cross section in the all-jet final state in $\Pp\Pp$ collisions at $\sqrt{s} = 13\TeV$}

\date{\today}

\abstract{
    A measurement of the production cross section of top quark pairs in association with two $\PQb$~jets ($\ttbar\bbbar$) is presented using data collected in proton-proton collisions at $\sqrt{s} = 13\TeV$ by the CMS detector at the LHC corresponding to an integrated luminosity of $35.9\fbinv$. The cross section is measured in the all-jet decay channel of the top quark pair by selecting events containing at least eight jets, of which at least two are identified as originating from the hadronization of $\PQb$~quarks. A combination of multivariate analysis techniques is used to reduce the large background  from multijet events not containing a top quark pair, and to help discriminate between jets originating from top quark decays and other additional jets. The cross section is determined for the total phase space to be $5.5 \pm 0.3\stat^{+1.6}_{-1.3}\syst\unit{pb}$ and also measured for two fiducial $\ttbar\bbbar$ definitions. The measured cross sections are found to be larger than theoretical predictions by a factor of 1.5--2.4, corresponding to 1--2 standard deviations.
}

\hypersetup{%
pdfauthor={CMS Collaboration},%
pdftitle={Measurement of the ttbb production cross section in the all-jet final state in pp collisions at sqrt(s) = 13 TeV},%
pdfsubject={CMS},%
pdfkeywords={CMS, physics, top, measurement}}

\newcommand{\ttbarbb}{\ensuremath{\ttbar\bbbar}\xspace}
\newcommand{\ttbarb}{\ensuremath{\ttbar\PQb}\xspace}
\newcommand{\ttbartwob}{\ensuremath{\ttbar2\PQb}\xspace}
\newcommand{\ttbarcc}{\ensuremath{\ttbar\PQc\PAQc}\xspace}
\newcommand{\ttbarlf}{\ensuremath{\ttbar\cmsSymbolFace{jj}}\xspace}

\newlength\cmsTabSkip\setlength{\cmsTabSkip}{1ex}

\maketitle

\section{Introduction}
\label{sec:intro}

At the CERN LHC, top quark pairs are produced with copious amounts of additional jets, including those resulting from the hadronization of $\PQb$ quarks ($\PQb$ jets).
Top quark pair production in association with a pair of $\PQb$ jets, $\ttbarbb$, is challenging to model because of the very different energy scales for the $\PQb$ jets produced in association with the $\ttbar$ system and that of $\ttbar$ system~\cite{Bredenstein:arXiv1001.4006}, and because of the small but nonnegligible mass of the \PQb quark.
Improving the accuracy and the precision of perturbative calculations in quantum chromodynamics (QCD) for this process is crucial, since it represents an important background for numerous searches or other measurements at the LHC.
In particular, $\ttbar$ production in association with a Higgs boson ($\ttbar\PH$), where the Higgs boson decays to $\bbbar$, suffers from an irreducible \ttbarbb background~\cite{Aaboud:2018urx,Sirunyan:2018hoz,Aaboud:2017rss,Aad:2016zqi,Sirunyan:2018ygk,Sirunyan:2018mvw}.
Searches for four top quark production ($\ttbar\ttbar$) are also affected by this background~\cite{Aaboud:2018jsj,Khachatryan:2014sca,Sirunyan:2017tep}.
The two latter processes provide direct access to the top quark Yukawa coupling, a crucial parameter of the standard model~\cite{Cao:2016wib,Cao:2019ygh}.
An improved understanding of the \ttbarbb process would help reduce the uncertainty in such measurements.

Calculations of the production cross section of \ttbar in association with jets have been performed at next-to-leading order (NLO) in QCD and matched with parton showers for up to two additional massless partons in the matrix element~\cite{Hoeche:2014qda,Alwall:2014hca,Frederix:2012ps}.
The \ttbarbb cross section at NLO, matched with parton showers, has also been calculated for massless $\PQb$ quarks (five-flavour scheme, 5FS)~\cite{Garzelli:2014aba}, and has recently become available for massive $\PQb$ quarks (four-flavour scheme, 4FS)~\cite{Cascioli:2013era,Bevilacqua:2017cru,Jezo:2018yaf}.
A comparison of the measurements of the \ttbarbb cross section with such calculations provides valuable guidance to improve the different frameworks.
The \ttbarbb cross section has been measured previously at $\sqrt{s}=8$ and $13\TeV$ by the ATLAS and CMS Collaborations, in events containing one or two charged leptons~\cite{Aad:2015yja,Aaboud:2018eki,CMS:2014yxa,Khachatryan:2015mva,Sirunyan:2017snr}.

This Letter focuses on the all-jet final state of the \ttbar system, where each top quark decays into three jets, leading to a signature of four $\PQb$ jets and four light-quark jets for the \ttbarbb system.
This final state is favoured by a large branching fraction and provides a complete reconstruction of top quarks, as opposed to other decay channels of the top quark pairs. Moreover, the main uncertainties affecting the  sensitivity in this measurement are different than those affecting final states containing leptons, therefore providing complementary information. However, the all-jet channel also suffers from a large background from multijet production, as well as from the difficulty of identifying jets that originate from decaying top quarks.
Multivariate analysis techniques are developed and implemented to mitigate these problems.
The \ttbarbb cross section is measured using data collected by the CMS detector in $\Pp\Pp$ collisions at $\sqrt{s}=13\TeV$, corresponding to an integrated luminosity of $35.9\fbinv$~\cite{CMS-PAS-LUM-17-001}.

\section{The CMS detector and event simulation}
\label{sec:cms_reco}

The central feature of the CMS apparatus is a superconducting solenoid of 6\unit{m} internal diameter, providing a magnetic field of 3.8\unit{T}. A silicon pixel and strip tracker, a lead tungstate crystal electromagnetic calorimeter (ECAL), and a brass and scintillator hadron calorimeter (HCAL), each composed of a barrel and two endcap sections reside within the solenoid field. Forward calorimeters extend the pseudorapidity coverage provided by the barrel and end detectors. Muons are detected in gas-ionization chambers embedded in the steel flux-return yoke outside the solenoid. A more detailed description of the CMS detector, together with a definition of its coordinate system and kinematic variables, can be found in Ref.~\cite{Chatrchyan:2008zzk}.
{\tolerance=8000
Samples of $\ttbar$ events are simulated at NLO in QCD using \POWHEG~(v2)~\cite{Nason:2004rx,Frixione:2007vw,Alioli:2010xd,Frixione:2007nw}.
These samples include $\ttbarbb$ events, where the additional \PQb jets are generated by the parton shower.
Single top quark production in the $t$ channel or in association with a $\PW$ boson, and $\ttbar\PH$ production are simulated at NLO with \POWHEG~\cite{Powheg_st,Re:2010bp,Hartanto:2015uka}.
Production of $\PW$ or $\PZ$ bosons in association with jets ($\PV$+jets), as well as QCD multijet events, are simulated at leading order (LO) with \MGvATNLO~(v2.2.2)~\cite{Alwall:2014hca}, and the MLM merging scheme~\cite{Alwall:2007fs}.
The \MGvATNLO generator is used at NLO for simulating associated production of top quark pairs with $\PW$ or $\PZ$ bosons ($\ttbar\PV$).
Diboson processes ($\PW\PW$, $\PW\PZ$ and $\PZ\PZ$) are simulated at LO using \PYTHIA~(v8.219)~\cite{Sjostrand:2014zea}.
\par}
All simulated events are processed with \PYTHIA for modelling of the parton showering, hadronization, and underlying event (UE).
The NNPDF~3.0~\cite{Ball:2014uwa} parton distribution functions (PDFs) are used throughout, at the same perturbative order as used by the event generators.
The CUETP8M1 UE tune~\cite{Khachatryan:2110213} is used for all processes except for the $\ttbar$, $\ttbar\PH$ and single top quark processes.
For these, an updated version of the tune is used (CUETP8M2T4), in which an adjusted value of the strong coupling constant is used in the description of initial-state radiation~\cite{CMS:2016kle}.
Simulation of the CMS detector response is based on \GEANTfour~(v9.4)~\cite{Agostinelli2003250}.
Additional $\Pp\Pp$ interactions in the same or neighbouring bunch crossings (pileup) are simulated with \PYTHIA and overlaid with hard-scattering events according to the pileup distribution measured in data.

The various simulated processes are normalized to state-of-the-art predictions for the production cross sections.
The $\ttbar$, $\PV$+jets, single top quark, and $\PW^+\PW^-$ samples are normalized to next-to-NLO (NNLO) precision in QCD~\cite{Czakon:2011xx,Kidonakis:2013zqa,Li:2012wna,Gehrmann:2014fva}, while remaining processes such as $\ttbar\PV$, $\ttbar\PH$, and other diboson production are normalized to NLO in QCD~\cite{Alwall:2014hca,Campbell:2011bn}.

\section{Definitions of fiducial phase space}
\label{sec:ttxx_def}

The $\ttbarbb$ production cross section is measured for three different phase space definitions.
Two definitions for \ttbarbb events in the fiducial phase space, matching the detector acceptance, are considered: one that is based exclusively on stable generated particles after hadronization (parton-independent), and one that also uses parton-level information after radiation emission (parton-based).
The former facilitates comparisons with predictions from event generators, while the latter is closer to the approach taken by searches for $\ttbar\PH$ production to define the contribution from the $\ttbarbb$ process.
The cross section is reported for the total phase space by correcting the parton-based fiducial cross section by the experimental acceptance.

Particle-level jets are defined by clustering stable generated final-state particles, excluding neutrinos, using the anti-\kt algorithm~\cite{Cacciari:2008gp, Cacciari:2011ma} with a distance parameter of 0.4.
These jets are defined unambiguously as $\PQb$ or $\PQc$ jets by rescaling the momenta of generated $\PQb$ and $\PQc$ hadrons to a negligible value, while preserving their direction, and including them in the clustering procedure~\cite{Cacciari:2007fd}.
A jet is labelled $\PQb$ jet if it is matched to at least one $\PQb$ hadron, and labelled $\PQc$ jet if matched with at least one $\PQc$ hadron and no $\PQb$ hadron.

Events in the generated $\ttbar$ sample are divided into exclusive categories according to the flavour of the jets that do not originate from the decay of top quarks, which we refer to as ``additional'' jets.
The $\PQb$ or $\PQc$ jets are considered to originate from a top quark if one of the clustered $\PQb$ or $\PQc$ hadrons features a top quark in its simulation history.
Additional jets are required to have a transverse momentum $\pt > 20\GeV$, and absolute pseudorapidity $\abs{\eta}<2.4$. No explicit requirement on the $\PQb$ hadron kinematic variables is used.
Events are categorized as $\ttbarbb$ if they contain at least two additional $\PQb$ jets, which defines the total phase space for which the \ttbarbb cross section is measured.
Events with a single additional $\PQb$ jet are categorized as $\ttbarb$ ($\ttbartwob$) if that $\PQb$ jet is matched with exactly one (at least two) $\PQb$ hadron(s).
The $\ttbarb$ events correspond to $\ttbarbb$ events where one of the additional $\PQb$ jets fails the above kinematic requirements, while $\ttbartwob$ events arise from collinear gluon splittings. If no $\PQb$ jets are present but at least one additional $\PQc$ jet is present the event is referred to as $\ttbarcc$; all remaining events are denoted $\ttbarlf$.

For the parton-based definition of the $\ttbarbb$ fiducial phase space, at least eight jets with $\pt > 20\GeV$ and $\abs{\eta}<2.4$ must be present, of which at least six have $\pt > 30\GeV$.
At least four of these jets must be $\PQb$ jets, and at least two of those must not originate from top quarks.
This last requirement is removed for the parton-independent fiducial definition, in order to be independent of the origin of the $\PQb$ jets, and thus of the simulated parton content.
Some \ttbarbb events in the total phase space failing the fiducial requirements may still be reconstructed and selected because of resolution effects, and are referred to as out-of-acceptance. They correspond to 16\% of all reconstructed \ttbarbb events.

\section{Event reconstruction and selection}
\label{sec:reco_sel}

The particle-flow algorithm~\cite{Sirunyan:2017ulk} aims to reconstruct and identify each particle in an event, with an optimized combination of information from the various elements of the CMS detector.
The primary $\Pp\Pp$ interaction vertex is taken to be the reconstructed vertex with the largest sum of the $\pt^2$ of the objects associated to that vertex, where the considered objects are those returned by a jet clustering algorithm~\cite{Cacciari:2008gp, Cacciari:2011ma} applied to the tracks assigned to the vertex, and the associated missing transverse momentum, taken as the negative vector sum of the \pt of those objects.
The energy of photons is obtained from the ECAL measurement. The energy of electrons is determined from a combination of the electron momentum at the primary interaction vertex as determined by the tracker, the energy of the corresponding ECAL cluster, and the energy sum of all bremsstrahlung photons spatially compatible with originating from the electron track.
The \pt of muons is obtained from the curvature of the corresponding tracks.
The energy of charged hadrons is determined from a combination of their momentum measured in the tracker and the matching ECAL and HCAL energy deposits, corrected for zero-suppression effects and for the response function of the calorimeters to hadronic showers.
The energy of neutral hadrons is obtained from the corresponding corrected ECAL and HCAL energies.

For each event, hadronic jets are clustered from the reconstructed particles using the anti-\kt algorithm with a distance parameter of 0.4. The jet momentum is determined as the vectorial sum of all particle momenta in the jet, and is found from simulation to be within 5 to 10\% of the true momentum over the whole \pt spectrum and detector acceptance.
Pileup interactions can contribute additional tracks and calorimetric energy depositions to the jet momentum. To mitigate this effect, tracks identified to be originating from pileup vertices are discarded and an offset correction is applied to correct for remaining contributions~\cite{Cacciari:2007fd}.
Jet energy corrections are derived from simulation to bring the average measured response of a jet to that of a particle-level jet. In situ measurements of the momentum balance in dijet, photon+jet, $\PZ$+jet, and multijet events are used to account for any residual differences in jet energy scale in data and simulation~\cite{Khachatryan:2016kdb}.
The data used for these measurements are independent of those used for the present Letter.

A combined secondary vertex $\PQb$ tagging algorithm (CSVv2) is used to identify jets originating from the hadronization of $\PQb$ quarks~\cite{Sirunyan:2017ezt}, with an efficiency for identifying $\PQb$ jets in simulated \ttbar events of about 65\%.
The misidentification probability is about 10 and 1\% for $\PQc$ and light-flavour jets, respectively, where the latter refers to jets originating from the hadronization of $\PQu$, $\PQd$, $\PQs$ quarks or gluons.
The distribution of the discriminator score for $\PQb$ and light-flavour jets in the simulation is calibrated to match the distribution measured in control samples of \ttbar events with exactly two leptons (electrons or muons) and two jets, and $\PZ$ bosons produced in association with jets where the $\PZ$ bosons decay to pairs of electrons or muons.
The calibration is achieved by reweighting events using scale factors that are parameterized by the jet flavour, $\pt$, $\abs{\eta}$, and $\PQb$ tagging discriminator score~\cite{Sirunyan:2017ezt}.

Data are collected using two triggers~\cite{Khachatryan:2016bia}, both requiring at least six jets with $\abs{\eta}<2.4$.
The first (second) trigger considers jets with $\pt > 40$ ($30$) \GeV, and requires that the jet scalar \pt sum, \HT, exceeds $450$ ($400$) \GeV and that at least one (two) of the jets is (are) $\PQb$ tagged.
The efficiency of these triggers is measured in simulation, as well as in a data control sample collected using independent single-muon triggers.
The trigger efficiency in simulation is corrected to match the efficiency observed in the data by reweighting events using scale factors defined as the ratio between the efficiencies in the data and simulation.
For events satisfying the preselection criteria detailed below, the trigger efficiency is above 95\%.

An offline preselection is applied to data and simulated events, by requiring the presence of at least six jets with $\pt > 40\GeV$ and $\abs{\eta}<2.4$, of which at least two are $\PQb$ tagged, and $\HT > 500\GeV$.
Additional jets in the events are considered if they satisfy the requirements $\pt > 30\GeV$ and $\abs{\eta}<2.4$.
Events are vetoed if they contain electrons or muons with $\pt > 15\GeV$ and $\abs{\eta}<2.4$ that satisfy highly efficient identification criteria~\cite{Sirunyan:2018fpa,Khachatryan:2015hwa} and are isolated from hadronic activity. About 20\% of the \ttbarbb events in the fiducial phase space pass the offline selection.

\section{Multivariate analysis}
\label{sec:MVA}

The final state considered in this analysis suffers from a large background from multijet production, as well as from the difficulty to identify which jets do not stem from top quark decays.
To address these challenges and improve the sensitivity to the \ttbarbb signal, several multivariate analysis tools have been employed.

The multijet background can be discriminated from \ttbar production by observing that the latter is expected to contain four light-quark jets from $\PW$ boson decays per event, whereas the former is enriched in gluon jets.
Gluon and quark jets are separated using a quark-gluon likelihood (QGL) variable, based on jet substructure observables~\cite{CMS-PAS-JME-13-002,CMS-DP-2016-070}.
Using the individual jet QGL values, the likelihood of an event to contain $N_{\PQq}$ light-quark jets and $N_{\Pg}$ gluon jets is defined as
\begin{linenomath*}
\begin{equation}
    L(N_{\PQq},N_{\Pg}) = \sum_{\text{perm}} \left( \prod_{k={i_1}}^{i_{N_{\PQq}}} f_{\PQq}(\zeta_{k}) \right) \left(\prod_{m={i_{N_{\PQq}+1}}}^{i_{N_{\PQq}+N_{\Pg}}}  f_{\Pg}(\zeta_{m}) \right),
\end{equation}
\end{linenomath*}
where the sums run over all possible assignments of $N_{\PQq}$ jets to quarks (indices $k$) and $N_{\Pg}$ jets to gluons (indices $m$), $\zeta_{i}$ is the QGL discriminant of the $i^\text{th}$ jet, and $f_{\PQq}$ and $f_{\Pg}$ are the probability densities for $\zeta_{i}$ under the hypothesis of ($\PQu$, $\PQd$, $\PQs$, or $\PQc$) quark or gluon origin, respectively.
When computing $L(N_{\PQq},N_{\Pg})$, $\PQb$-tagged jets are not considered.
Based on the event likelihoods with $N_{\PQq}=4$ and $N_{\Pg}=0$, as well as $N_{\PQq}=0$ and $N_{\Pg}=4$, the QGL ratio (QGLR) is defined as $\text{QGLR} = L(4, 0)/(L(4, 0) + L(0, 4)).$
Other values for $N_{\PQq}$ and $N_{\Pg}$ have been tried but led to reduced discrimination between multijet and \ttbar production.
We correct the modelling of the QGL in the simulation by reweighting each event based on the quark or gluon origin and the QGL value of all jets in the event, where the weights are measured using data samples enriched in $\PZ$+jets and dijet events~\cite{CMS-DP-2016-070}.
After applying this correction, a good agreement is found between data and simulation.

To address the large combinatorial ambiguity in identifying the additional jets in the events, we have trained a boosted decision tree (BDT) using the TMVA package~\cite{Voss:1116810}, henceforth referred to as the ``permutation BDT''.
In events with eight reconstructed jets, there are 28 ways to select six of those as originating from the all-jet decay of a top quark pair, and there are 90 ways to match those six jets to the six partons from the top quark decay chains.
Some permutations are indistinguishable and are not considered, \ie permutations of two jets assigned to a $\PW$ boson decay are not considered, and neither are the permutations of three jets assigned to a $\PQt$ or $\PAQt$ decay.
To reduce the large number of permutations, the least favoured ones are rejected using a $\chi^2$ variable quantifying the compatibility of the invariant masses of the different jet pairings with those of the particles they should come from, defined as
\begin{linenomath}
\ifthenelse{\boolean{cms@external}}
{
\begin{multline*}
\chi^{2} = (m_{\mathrm{j}_1,\mathrm{j}_3,\mathrm{j}_4}-m_{\PQt})^{2}/\sigma_{\PQt}^{2} + (m_{\mathrm{j}_3,\mathrm{j}_4}-m_{\PW})^{2}/\sigma_{\PW}^{2} +  \\
(m_{\mathrm{j}_2,\mathrm{j}_5,\mathrm{j}_6}-m_{\PQt})^{2}/\sigma_{\PQt}^{2} + (m_{\mathrm{j}_5,\mathrm{j}_6}-m_{\PW})^{2}/\sigma_{\PW}^{2},
\label{eq:chi2}
\end{multline*}
} 
{ 
\begin{equation*}
\chi^{2} = (m_{\mathrm{j}_1,\mathrm{j}_3,\mathrm{j}_4}-m_{\PQt})^{2}/\sigma_{\PQt}^{2} + (m_{\mathrm{j}_3,\mathrm{j}_4}-m_{\PW})^{2}/\sigma_{\PW}^{2} +  (m_{\mathrm{j}_2,\mathrm{j}_5,\mathrm{j}_6}-m_{\PQt})^{2}/\sigma_{\PQt}^{2} + (m_{\mathrm{j}_5,\mathrm{j}_6}-m_{\PW})^{2}/\sigma_{\PW}^{2},
\label{eq:chi2}
\end{equation*}
}
\end{linenomath}
where $m_{(\dots)}$ denotes the invariant mass of the given jets, and $\sigma_{\PW}=10.9\GeV$ and $\sigma_{\PQt}=17.8\GeV$ are the experimental resolutions in the two- and three-jet invariant masses, respectively.
The masses entering the equation are $m_{\PQt} = 172.3\GeV$ and $m_{\PW} = 80.2\GeV$, measured from the generated \ttbar system after reconstruction.
The BDT is trained using simulated \ttbar events after applying the above preselection criteria, requiring the presence of at least seven jets, and reducing the number of permutations by requiring that $\chi^{2} < 33.38$, corresponding to a p-value $P(\chi^2)$ of $10^{-6}$ for a $\chi^{2}$ distribution with four degrees of freedom.
Events for which no permutation satisfies this requirement are rejected.
The correct jet-parton assignment is considered as a signal in the training, while all other distinguishable combinations are treated as background.
Input variables used for the BDT include jet $\PQb$ tagging discriminator scores and kinematic quantities, such as invariant masses of pairs and triplets of jets, angular openings between jets, and the transverse momenta of jets.
For each permutation, only quantities pertaining to the six jets assumed to originate from the top quarks are used in the training.
The permutation yielding the highest BDT score is used for the rest of the analysis.
For \ttbar events with eight jets where all six jets from the top quark decays have been selected, the permutation BDT identifies the correct permutation with about 60\% efficiency.

As a further handle to reduce the multijet background, we have trained a second BDT to discriminate this background from inclusive $\ttbar$+jets production.
While supervised training of multivariate classifiers relies on samples of simulated events, the poor modelling of multijet production and the insufficient size of the available simulated samples limit the achievable discrimination power.
A proposed method to alleviate these shortcomings is a classification without labels (CWoLa)~\cite{Metodiev:2017vrx}.
In this weakly supervised approach, the classifier is trained using data, whereby one region in the data is treated as background and another independent region is treated as signal.
In the limit of large training sample the resulting classifier converges to the optimal classifier to distinguish between signal and  background, provided the two following conditions are fulfilled~\cite{Metodiev:2017vrx}.
First, the relative rates of the actual signal and background processes should be different in the two regions.
Second, the distributions of the variables entering the CWoLa classifier should be independent of the quantity used to define the two regions, for both the signal and background processes.
The CWoLa BDT is trained using a sample of data with exactly seven jets, where two independent regions are defined by requiring that the QGLR is below or above $0.95$.
The first and second regions are expected to contain about 10 and 20\% of $\ttbar$ events, respectively.
Variables used for constructing the CWoLa BDT are kinematic quantities similar to those used in the permutation BDT, the output value of the permutation BDT, and the \PQb tagging discriminator scores of the two jets identified by the permutation BDT as the $\PQb$ jets originating from the top quark decays.
Only the six jets identified by the permutation BDT as coming from the top quark decays are used to define the CWoLa BDT input variables.
The performance of the resulting classifier, measured in the region with at least eight jets, is found to be comparable to that of a supervised classifier trained using simulated samples.

\section{Cross sections}
\label{sec:xs_meas}

\begin{figure}[hbtp]
    \centering
    \includegraphics[width=0.45\textwidth]{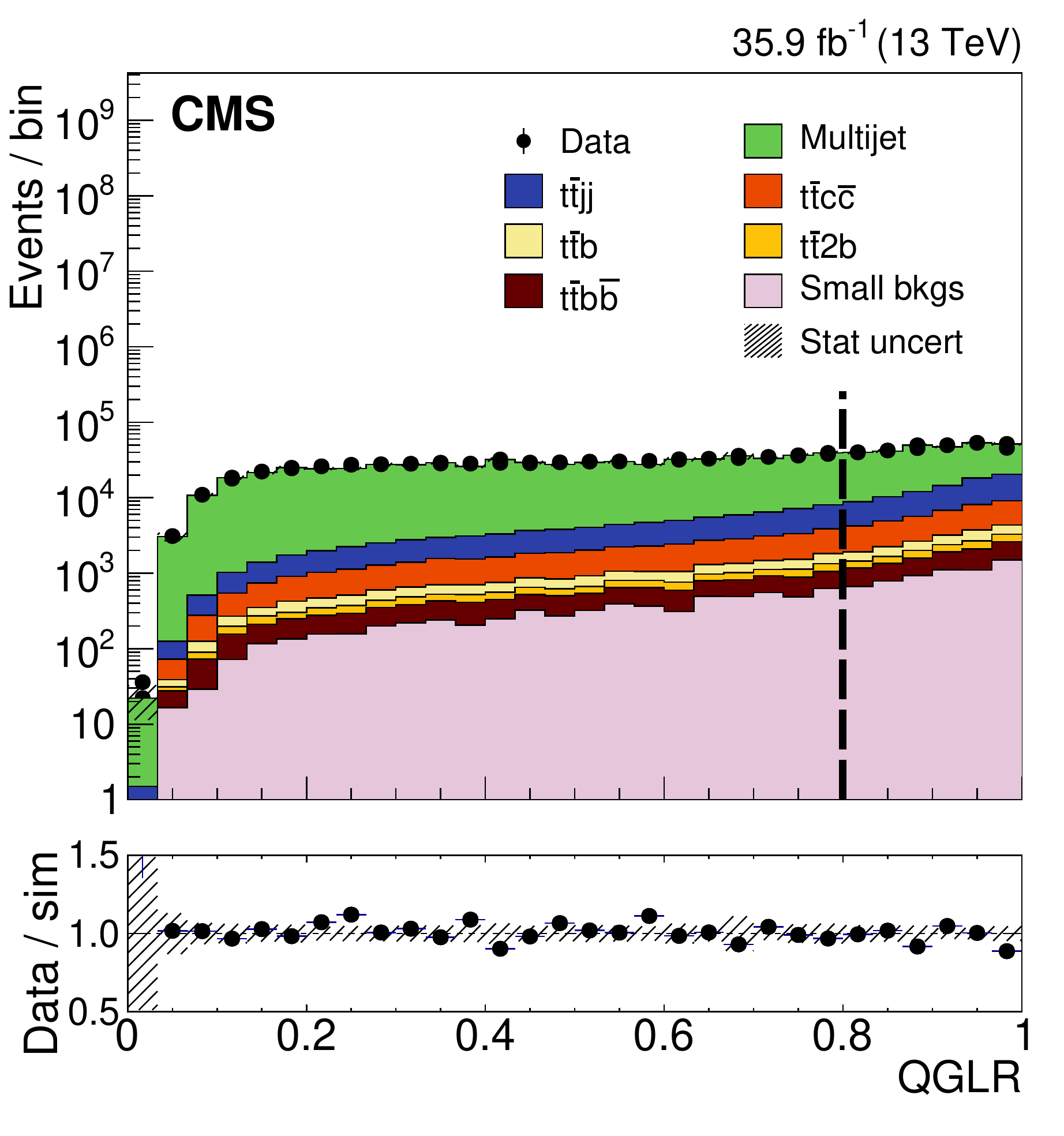}
    \includegraphics[width=0.45\textwidth]{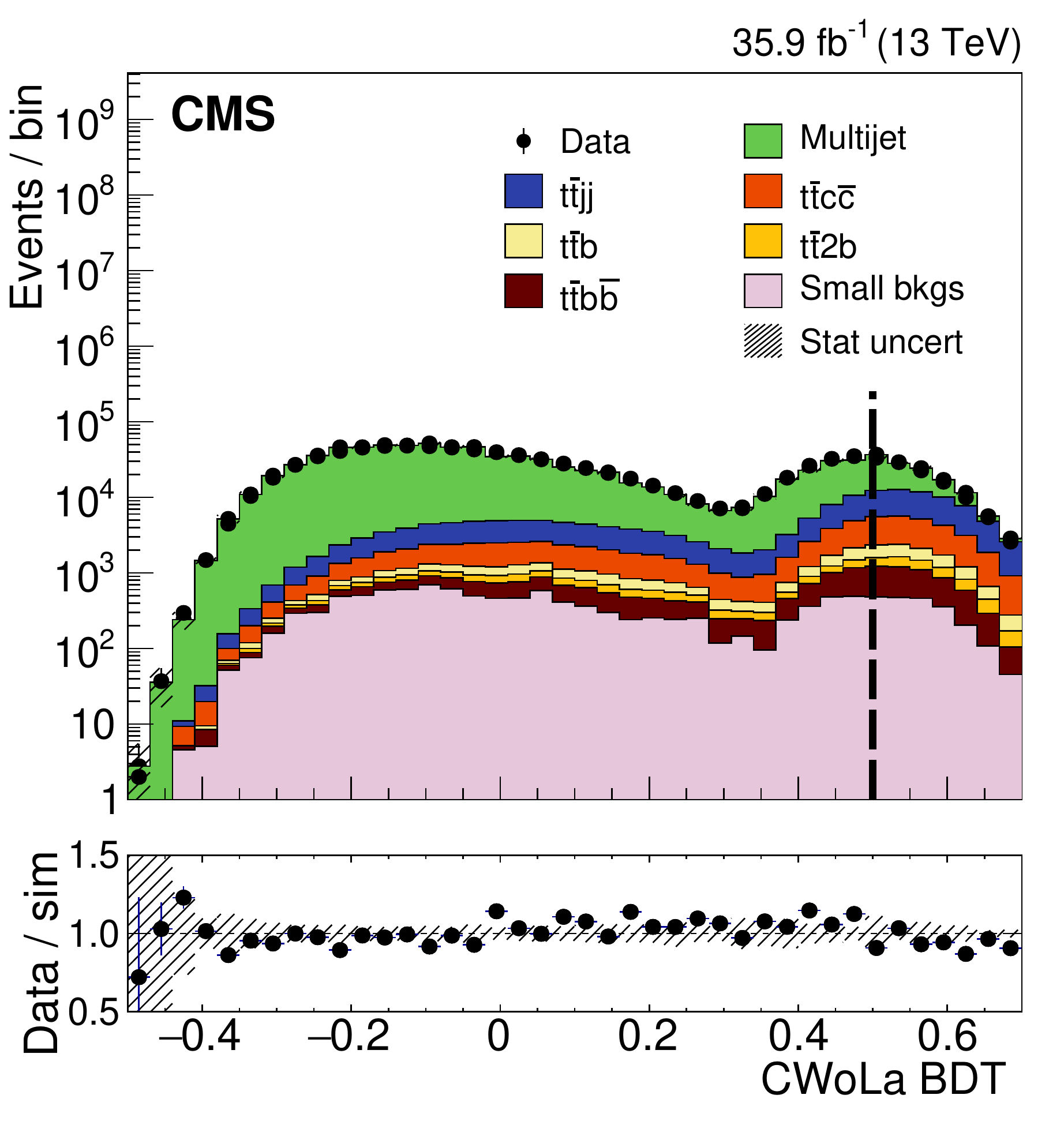}
    \caption{Distributions in the QGLR (\cmsLeft) and the CWoLa BDT discriminants (\cmsRight). Both are after preselection, requiring $P(\chi^2) > 10^{-6}$ and at least eight selected jets.
        All the contributions are based on simulation. The multijet contribution is scaled to match the total yields in data, after the other processes including the \ttbarbb signal have been normalized to their corresponding theoretical cross sections. This choice takes into account only the effect of the shape variation from the multijet background.
        The small backgrounds include $\ttbar\PV$, $\ttbar\PH$, single top quark, $\PV$+jets, and diboson production.
        The lower panels show the ratio between the observed data and the predictions.
        The dashed lines indicate the boundaries between the signal and control regions defined in Section \ref{sec:xs_meas}.
        Hatched bands indicate the statistical uncertainty in the predictions without considering the systematic sources, dominated by the uncertainties in the simulated multijet background. Underflow and overflow events were added to the first and last bins, respectively.
    }
    \label{fig:qglr_cwola}
\end{figure}

To measure the \ttbarbb cross section we require, in addition to the preselection criteria, the presence of at least eight jets, and $P(\chi^2) > 10^{-6}$.
The distributions in the QGLR and of the CWoLa BDT discriminants for selected events are shown in Fig.~\ref{fig:qglr_cwola}.
The cross section is extracted from a binned maximum likelihood fit to a two-dimensional distribution (referred to as 2DCSV) constructed using the largest and second-largest $\PQb$ tagging discriminator scores among the jets determined to be additional jets by the permutation BDT.
In order to increase the signal purity and the precision in the measurement, we define a signal region (SR) by requiring that the CWoLa BDT score be above 0.5, and the QGLR be above 0.8.
These thresholds are optimized to obtain the best expected precision in the cross section.
About 20\% of the \ttbarbb signal that passes the offline preselection is selected into the SR.

The multijet background is also estimated from data.
Three independent control regions (CRs), orthogonal to the SR, are defined by inverting the requirements on the CWoLa BDT and the QGLR: the CR1 ($\text{BDT}>0.5$, $\text{QGLR}<0.8$), the CR2 ($\text{BDT}<0.5$, $\text{QGLR}<0.8$), and the CR3 ($\text{BDT}<0.5$, $\text{QGLR}>0.8$).
For multijet production, the CWoLa BDT score and the QGLR are nearly independent, so that in each bin $i$ of the 2DCSV distribution the number of multijet events in the SR, $N^{\text{SR}}_i$, can be estimated from the number of multijet events in the CRs as
\begin{linenomath*}
\begin{equation}
    N^{\text{SR}}_i = N^{\text{CR3}}_i \, \frac{N^{\text{CR1}}_i}{N^{\text{CR2}}_i}. \label{eq:abcd}
\end{equation}
\end{linenomath*}
This relationship is a consequence of the choice of variables entering the CWoLa BDT, which were required to be independent of the QGLR in order to satisfy the hypotheses of the CWoLa method.
In order to properly take into account the small but non-negligible signal contribution in the CRs, the fit to extract the cross section is performed in all four regions, with the multijet rates $N^{\text{CR1}}_i$, $N^{\text{CR2}}_i$, and $N^{\text{CR3}}_i$ free to vary in the fit.
The assumption of Eq.~(\ref{eq:abcd}) on which this estimation relies is confirmed using the simulation.
In addition, we verify that Eq.~(\ref{eq:abcd}) is also satisfied in the data for kinematic distributions, such as the invariant mass of the reconstructed $\PW$ bosons and top quarks, where for each bin of these distributions the multijet yields are estimated by taking the difference between the observed yields in data and the predicted yields of all simulated processes.
Finally, we validate Eq.~(\ref{eq:abcd}) using alternative definitions of the four regions in the plane formed by the QGLR and the  CWoLa BDT, excluding the SR as defined above. The outcome of goodness-of-fit tests of the 2DCSV distribution was also positive for each of the alternative region definitions.

The data are fitted using a profiled maximum likelihood technique, where the likelihood is built as a product of independent Poisson likelihoods, defined for each bin $i$ of the 2DCSV distributions in the four event regions using the following expression for the number of events in bin $i$:
\begin{linenomath*}
\begin{equation}
    \mathcal{N}_i = \mu \, \mathcal{T}_i^{\text{sig}}(\vec{\theta}) + \sum_{k  \text{ in sim bkg}}  \mathcal{T}_{i}^k(\vec{\theta}) + N_i,
    \label{eq:fit}
\end{equation}
\end{linenomath*}
where $\mu$ is a signal strength parameter, defined by the ratio of observed to expected signal, $\mathcal{T}_i^{k}$ is the expected yield for process $k$ in bin $i$, ``sig'' includes the contributions from \ttbarbb, \ttbartwob, and \ttbarb, and $\vec{\theta}$ is a vector of nuisance parameters affecting the predicted yields of the various processes introduced to model the systematic uncertainties described in the next section.
The parameters $N_i$ are used to estimate the multijet background from the combined fit of the four regions; they are free parameters in the CRs and are given by Eq.~(\ref{eq:abcd}) in the SR.
The likelihood also features constraint terms for each of the nuisance parameters considered in the fit.
Different templates are constructed from \ttbarbb events matching the fiducial requirements and from events failing these requirements.
For the fiducial \ttbarbb templates, the effect of nuisance parameters corresponding to theoretical uncertainties is normalized such that the \ttbarbb cross section in the fiducial phase space is preserved, i.e. only shape variations within that phase space and their impact on the reconstruction efficiency are taken into account.
No such requirement is made for the other templates.
The uncertainty in the measured cross section is obtained by profiling the nuisance parameters.
As described in the next section, some uncertainties are not profiled and are added in quadrature with the uncertainty obtained from the fit.
The fit is repeated for each of the two fiducial phase-space definitions for \ttbarbb events described in Section~\ref{sec:ttxx_def}, leading to different in- and out-of-acceptance \ttbarbb templates.
The total \ttbarbb cross section is obtained by dividing the cross section for the parton-based fiducial phase space by the acceptance, estimated using {\POWHEG}+{\PYTHIA} to be $(29.4 \pm 1.8)\%$. Uncertainties affecting this acceptance correction are detailed in the next section.

\section{Systematic uncertainties}
\label{sec:syst}

Several sources of systematic uncertainties affecting the predictions for the signal and background processes entering the analysis are considered. These uncertainties may affect the normalization of the templates entering the fit, or may alter both their shape and their normalization.
The migration of events between the four regions is taken into account when relevant.
Experimental sources of uncertainties are taken to be fully correlated for all signal and background distributions estimated using the simulation, while only a subset of theoretical uncertainties are correlated among the $\ttbar$+jets components.

The modelling of the shape of the $\PQb$ tagging discriminator in the simulation represents an important source of systematic uncertainty.
Several uncertainties in the calibration of the $\PQb$ tagging discriminator distribution are propagated independently to the shape and normalization of the 2DCSV templates.
These are related to the uncertainty in the contamination by light- (heavy-) flavour jets in the control samples used for the measurement of heavy-  (light-) jet correction factors, as well as to the statistical uncertainty in these measurements~\cite{Sirunyan:2017ezt}.
Since no dedicated measurement is performed for $\PQc$ jets, the uncertainty in the shape of the $\PQb$ tagging discriminator distribution for $\PQc$ jets is conservatively taken to be twice the relative uncertainty considered for $\PQb$ jets.
In total, six different nuisance parameters are introduced to estimate the uncertainty arising from $\PQb$ tagging.

We evaluate the effect of the uncertainty in the jet energy scale (JES) and jet energy resolution (JER) by shifting the jet four-momenta using correction factors that depend on jet \pt and $\abs{\eta}$ for the JES, and jet $\abs{\eta}$ for the JER~\cite{Khachatryan:2016kdb}.
The calibration of the JES is affected by several sources of uncertainty, which are propagated independently to the measurement.
The uncertainty in the JES is also propagated to the $\PQb$ tagging calibration, and the resulting effect on the distribution of the $\PQb$ tagging discriminators is taken to be correlated with the effect on the jet momenta.

Uncertainties pertaining to the QGL are estimated conservatively by removing or doubling the scale factors applied to correct the distribution of the QGL in the simulation~\cite{CMS-DP-2016-070}.
The uncertainty in the integrated luminosity is evaluated to be 2.5\%~\cite{CMS-PAS-LUM-17-001}.
Uncertainties in the trigger efficiency are estimated by varying the trigger scale factors by their uncertainty, as determined from the efficiency measurements in data and simulation.
The uncertainty in the modelling of pileup is estimated by reweighting simulated events to yield different distributions of the expected number of pileup interactions, obtained by varying the total inelastic $\Pp\Pp$ cross section by 4.6\%~\cite{Aaboud:2016mmw}.
We take into account the limited size of the simulated samples by varying independently the predicted yields in every bin by their statistical uncertainties.

Theoretical uncertainties in the modelling of the $\ttbar$+jets process enter this analysis both through the efficiency to reconstruct and select \ttbarbb events, and through the contamination from \ttbarcc and \ttbarlf backgrounds.
The uncertainties in the renormalization and factorization scales ($\mu_\mathrm{R}$ and $\mu_\mathrm{F}$, respectively) are estimated by varying both scales independently by a factor of two up or down in the event generation, omitting the two cases where the scales are varied in opposite directions, and taking the envelope of the six resulting variations.
Likewise, the uncertainties related to the choice of the scale in the parton shower is evaluated by varying the scale in the initial-state shower by factors of 0.5 and 2, and the scale in the final-state shower by factors of $\sqrt{2}$ and $1/\sqrt{2}$.
Propagation of the uncertainties associated with the PDFs, as well as with the value of the strong coupling in the PDFs, has been achieved by reweighting generated events using variations of the NNPDF 3.0 set~\cite{Ball:2014uwa}.
The impact of the choice of the matching scale $h_{\text{damp}}=1.58 m_{\PQt}$ between the matrix-element generator and the parton shower in \POWHEG is evaluated using simulated samples generated with different choices of $h_{\text{damp}}=m_{\PQt}$ and $2.24 m_{\PQt}$~\cite{CMS:2016kle}.
We evaluate the uncertainty related to the UE tune by varying the tune parameters according to their uncertainties.
The uncertainty from the modelling of colour reconnection in the final state is evaluated by considering four alternatives to the \PYTHIA default, which is based on multiple-parton interactions (MPI) with early resonance decays (ERD) switched off.
These alternatives are an MPI-based scheme with ERD switched on, a QCD-inspired scheme~\cite{Christiansen:2015yqa}, and a gluon-move scheme with ERD either off or on~\cite{Argyropoulos:2014zoa}. All the alternative models were tuned to LHC data \cite{Sirunyan:2018avv}.
It has been verified that the selection efficiency obtained from the nominal $\ttbar$ simulation, in which additional \PQb jets are generated by the parton shower, is in agreement within estimated modelling uncertainties with that obtained using a sample of \ttbarbb events generated at NLO in QCD with massive \PQb quarks (4FS)~\cite{Jezo:2018yaf}.
Since the spectrum of the top quark $\pt$ is known to be softer in the data than in the simulation, we evaluate the effect of this mismodelling by reweighting the generated events to match the top quark $\pt$ distribution measured in data~\cite{Sirunyan:2018wem}.
The latter two uncertainties are not evaluated using profiled nuisance parameters, but by repeating the measurement using varied signal and background predictions.
The differences in the measured cross sections are taken as the corresponding uncertainties and are added in quadrature with the uncertainty obtained from the profile likelihood.
Uncertainties related to the $\mu_\mathrm{R}$ and $\mu_\mathrm{F}$ scales, the parton shower scale, and the $h_{\text{damp}}$ choice are taken to be uncorrelated for the \ttbarbb, \ttbarb, \ttbartwob,  \ttbarcc and \ttbarlf templates, while the other modelling uncertainties are taken to be correlated for all \ttbar events.
In addition to the aforementioned modelling uncertainties, we assign an uncertainty of 50\% to the normalization of the \ttbarcc background to cover the lack of precise measurements of this process.
The results are stable when doubling that uncertainty.

Compared to $\ttbar$+jets and multijet production, the contribution of other background processes such as $\ttbar\PV$, $\ttbar\PH$, $\PV$+jets, diboson, and single top quark production is small.
We assign uncertainties to their predicted rates based on the PDF and $\mu_\mathrm{R}$/$\mu_\mathrm{F}$ scale uncertainties in their theoretical cross sections.

Table~\ref{tab:syst_sources} summarizes the contributions of the various sources of systematic uncertainties to the total uncertainty in the cross sections measured in the fiducial phase space.
The theoretical uncertainty in the acceptance from the various sources listed above is estimated to be 6\%, and is added in quadrature with the uncertainty in the parton-based fiducial cross section to yield the systematic uncertainty in the total \ttbarbb cross section.

\begin{table*}[htb]
    \renewcommand{\arraystretch}{1.5}
    \centering
    \topcaption{The considered sources of systematic uncertainties and their respective contributions to the total systematic uncertainty in the measured \ttbarbb cross section for the two defined \ttbarbb fiducial phase spaces.
    The upper (lower) portion of the table lists uncertainties related to the experimental conditions (theoretical modelling). The numbers are obtained by taking the difference in quadrature of the profile likelihood width when fixing  nuisance parameters corresponding to a given source of uncertainty and leaving the others free to vary.}
    \label{tab:syst_sources}

    \begin{tabular}{lcc}
        Source     & \shortstack{Fiducial, \\ parton-independent (\%)} & \shortstack{Fiducial, \\ parton-based (\%)}  \\
        \hline
        Simulated sample size                 & ${}^{+15}_{-11}$ & ${}^{+15}_{-11}$ \\
        Quark-gluon likelihood                & ${}^{+13}_{-8}$  & ${}^{+13}_{-8}$  \\
        \PQb tagging of \PQb quark                          & $\pm 10$  & $\pm 10$  \\
        JES and JER                            & ${}^{+5.1}_{-5.2}$   & ${}^{+5.0}_{-5.4}$   \\
        Integrated luminosity                 & ${}^{+2.8}_{-2.2}$   & ${}^{+2.4}_{-2.2}$   \\
        Trigger efficiency                    & ${}^{+2.6}_{-2.1}$   & ${}^{+2.5}_{-2.2}$   \\
        Pileup                                & ${}^{+2.3}_{-2.0}$   & ${}^{+2.2}_{-1.9}$   \\
        \hline
        $\mu_{\mathrm{R}}$ and $\mu_{\mathrm{F}}$ scales        & ${}^{+13}_{-9}$  & ${}^{+13}_{-9}$ \\
        Parton shower scale                   & ${}^{+11}_{-8}$  & ${}^{+11}_{-8}$ \\
        UE tune                               & ${}^{+9.0}_{-5.3}$  & ${}^{+9.0}_{-5.2}$   \\
        Colour reconnection                   & $\pm 7.2$  & $\pm 7.1$\\
        Shower matching ($h_{\text{damp}}$)   & ${}^{+4.3}_{-2.8}$  & ${}^{+3.8}_{-2.7}$   \\
        \ttbarcc normalization                & ${}^{+3.2}_{-4.4}$  & ${}^{+2.9}_{-4.5}$   \\
        Modelling of \pt of top quark               & $\pm 2.5$ & $\pm 2.4$ \\
        PDFs                                  & ${}^{+2.2}_{-2.0}$   & ${}^{+2.2}_{-2.0}$   \\
        \hline
        Total                                 & ${}^{+28}_{-23}$   & ${}^{+28}_{-23}$ \\
    \end{tabular}
\end{table*}

\section{Results}
\label{sec:results}

\begin{figure*}[hbtp]
    \centering
    \includegraphics[width=0.45\textwidth]{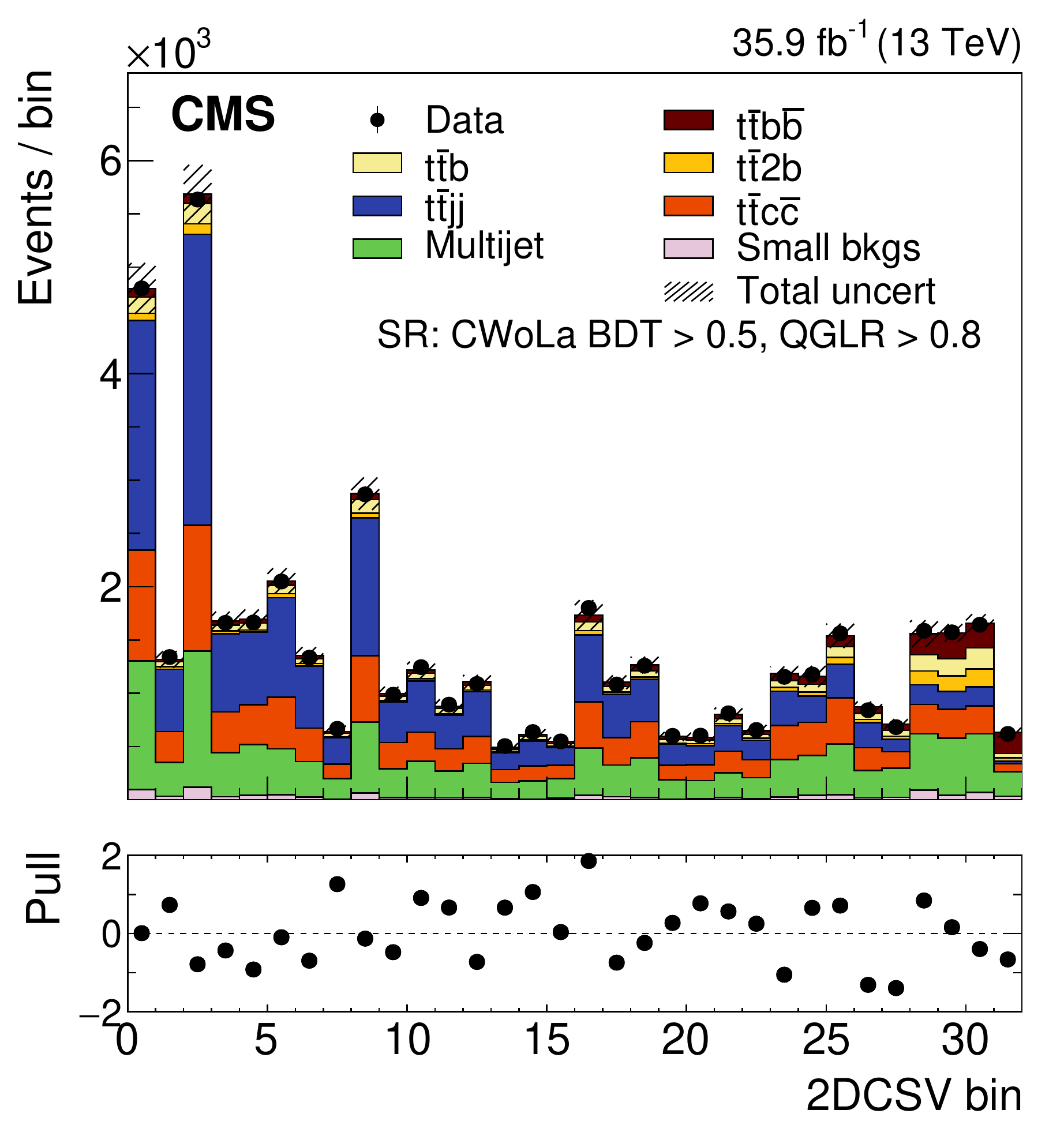}
    \includegraphics[width=0.45\textwidth]{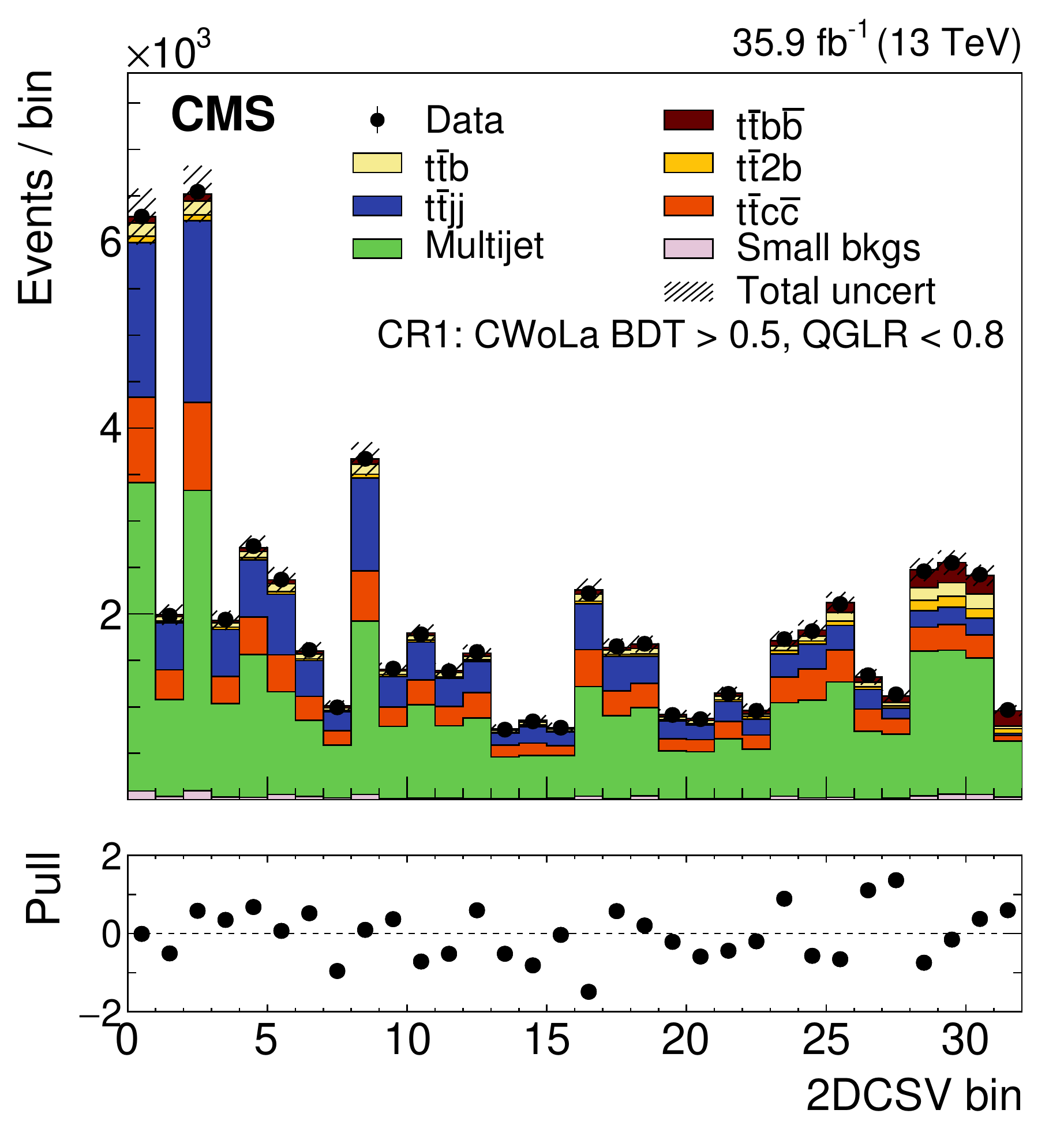}
    \includegraphics[width=0.45\textwidth]{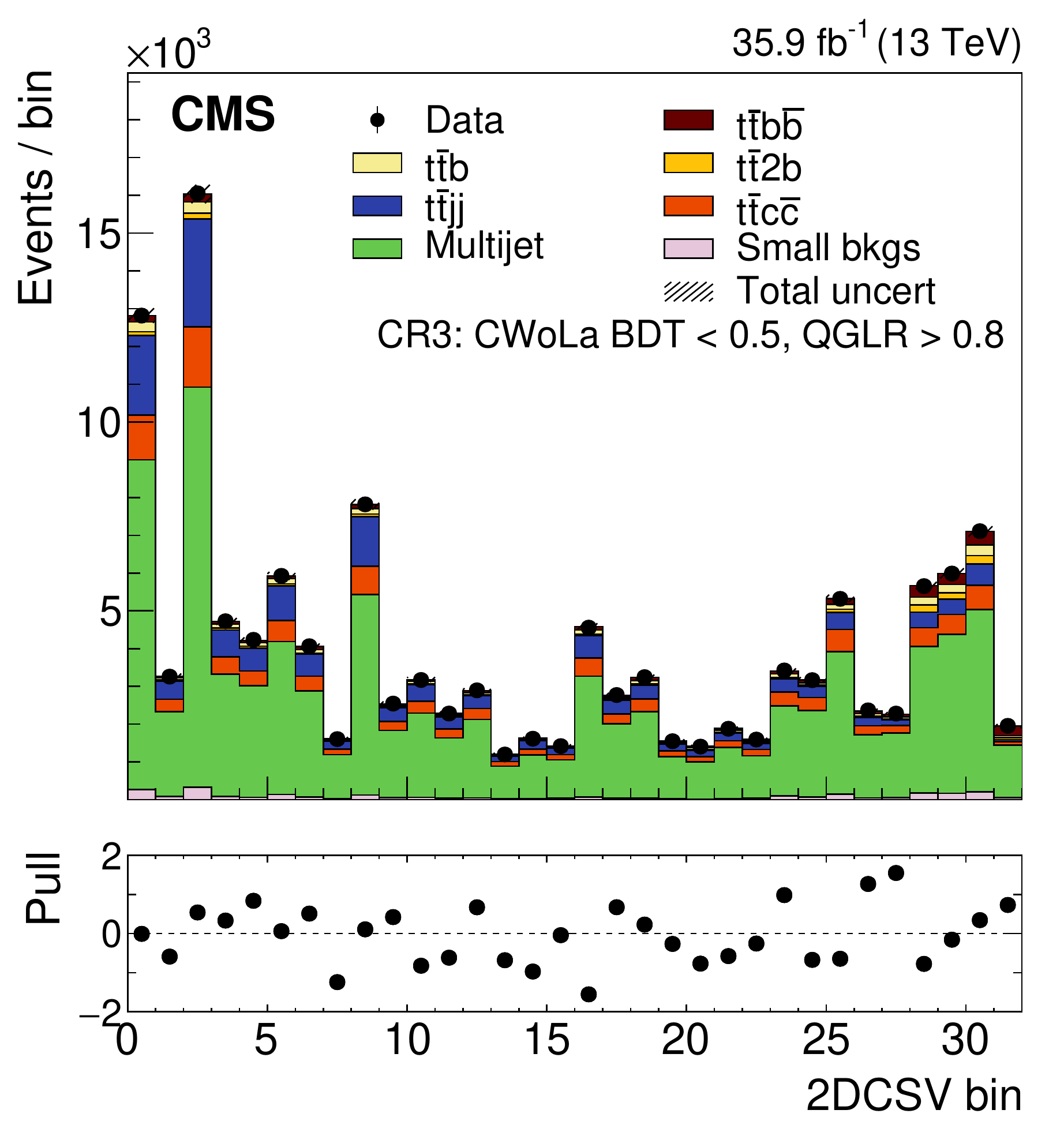}
    \includegraphics[width=0.45\textwidth]{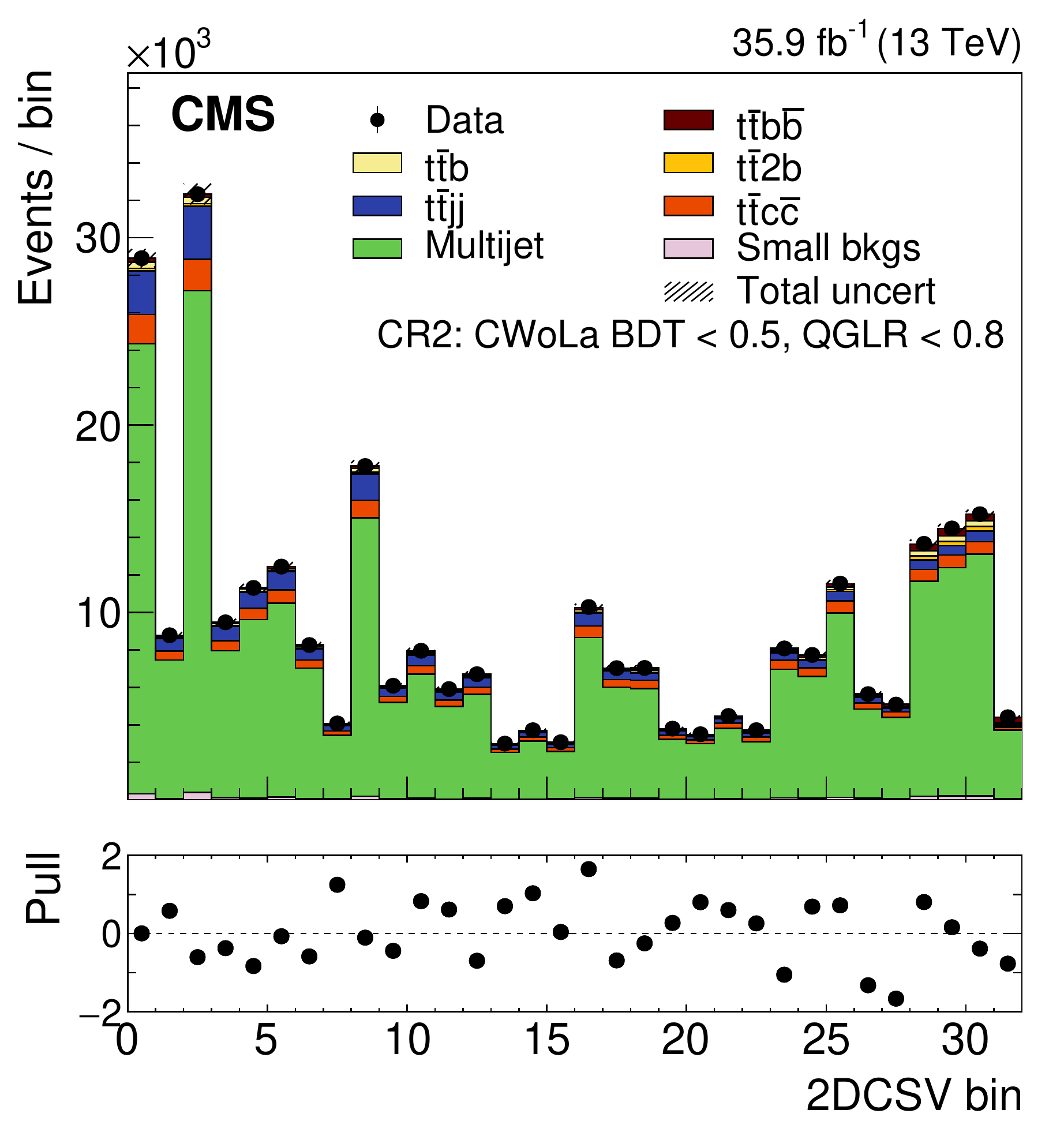}
    \caption{
        Distribution in the 2DCSV in the SR (upper left), CR1 (upper right), CR2 (lower right), and CR3 (lower left) regions.
        For clarity, the two-dimensional distribution with largest and next-to-largest \PQb tagging discriminant scores for the additional jets have been unrolled to one dimension, and the resulting bins  ordered according to increasing values of the ratio between expected signal and background yields in each bin of the SR.
        The small backgrounds include $\ttbar\PV$, $\ttbar\PH$, single top quark, $\PV$+jets, and diboson production.
        Hatched bands correspond to uncertainties. The bottom panels show the pull distribution. The pull is defined as the bin by bin difference between data and predicted yields after the fit, divided by the uncertainties accounted for correlations between data and predictions after the fit.}
    \label{fig:postfit}
\end{figure*}

The result of the maximum likelihood fit described in Section~\ref{sec:xs_meas} is shown in Fig.~\ref{fig:postfit} for the 2DCSV distributions in the four analysis regions.
The contribution from multijet production nearly matches the differences between the yields in data and from the other processes in the CR1, CR2, and CR3 because it is estimated from the data in the four regions according to the method described in the previous section.
The measured cross section for the two \ttbarbb definitions in the fiducial phase space, as well as for the total phase space introduced in Section~\ref{sec:ttxx_def}, are given in Table~\ref{tab:results}. The measurement uncertainty is dominated by the systematic effects from the simulation sample sizes, QGL corrections, and $\mu_\mathrm{R}$ and $\mu_\mathrm{F}$ dependences on changes in scale.

\begin{figure*}[hbtp]
    \centering
    \includegraphics[width=0.95\textwidth]{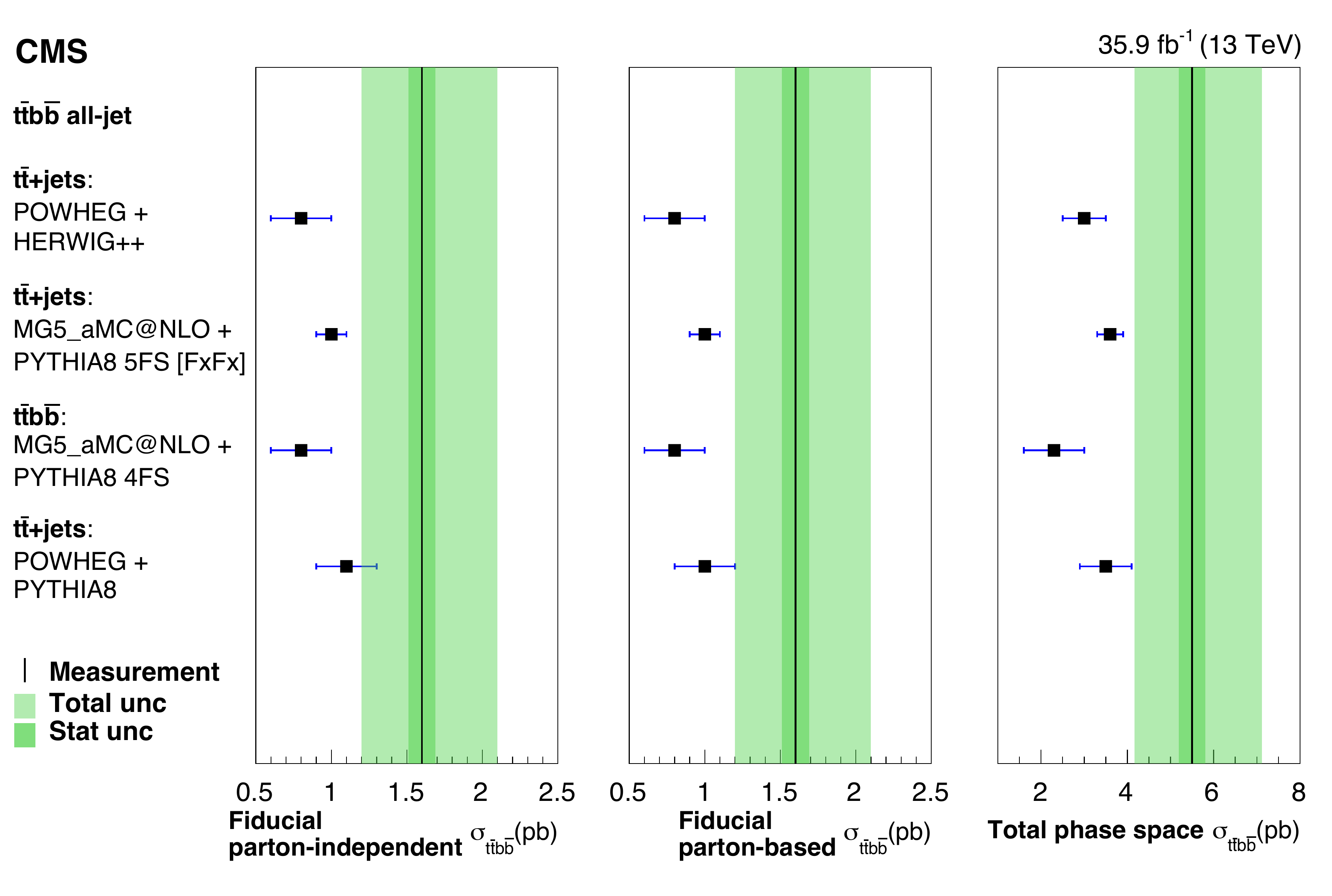}
    \caption{
        Comparison of the measured \ttbarbb production cross sections (vertical lines) with predictions from several Monte Carlo generators (squares), for three definitions of our \ttbarbb regions of phase space: fiducial parton-independent (left), fiducial parton-based (middle), total (right). The dark (light) shaded bands show the statistical (total) uncertainties in the measured value.
        Uncertainty intervals in the theoretical cross sections include the statistical uncertainty as well as the uncertainties in the PDFs and the $\mu_\mathrm{R}$ and $\mu_\mathrm{F}$ scales.
        }
    \label{fig:results}
\end{figure*}

Because of the large overlap between the two definitions of the \ttbarbb fiducial phase space, the measured cross sections are numerically equal at the quoted precision.
The measurements are compared with NLO predictions from \POWHEG for inclusive \ttbar production interfaced with either \PYTHIA or \HERWIGpp~(v2.7.1)~\cite{Bahr:2008pv}, using the EE5C UE tune~\cite{Gieseke:2012ft} for the latter.
Predictions from \MGvATNLO at NLO interfaced with \PYTHIA for \ttbar production with up to two extra massless partons (5FS) merged using the FxFx scheme~\cite{Frederix:2012ps}, and for \ttbarbb production with massive $\PQb$ quarks (4FS), are also compared with the measurements.
The predicted cross sections are not rescaled by any NLO to NNLO K-factor, which for inclusive \ttbar production amounts to 1.1--1.15~\cite{Czakon:2011xx}.
Measured and predicted cross sections are shown in Fig.~\ref{fig:results}.
The predictions underestimate the measured cross section by a factor of 1.5--2.4, corresponding to differences of 1--2 standard deviations. This is consistent with the results from Refs.~\cite{Aad:2015yja,Aaboud:2018eki,CMS:2014yxa,Khachatryan:2015mva,Sirunyan:2017snr}.

\begin{table*}[htb]
    \renewcommand{\arraystretch}{1.5}
    \centering
    \topcaption{Measured and predicted cross sections for the different definitions of the \ttbarbb phase space considered in this analysis.
        For measurements, the first uncertainty is statistical, while the second one is from the systematic sources.
        The uncertainties in the predicted cross sections include the statistical uncertainty, the PDF uncertainties, and the $\mu_\mathrm{R}$ and $\mu_\mathrm{F}$ dependences on changes in scale.
        The uncertainties in scale for parton showers are not included, and amount to about 15\% for {\POWHEG}+{\PYTHIA}.
        Unless specified otherwise, \PYTHIA is used for the modelling the parton shower, hadronization, and the underlying event.
        \label{tab:results}}

    \begin{tabular}{lccc}
        & \shortstack{Fiducial, \\ parton-independent (pb)} & \shortstack{Fiducial, \\ parton-based (pb)} & Total (pb) \\
        \hline
        Measurement & $1.6 \pm 0.1 ^{+0.5}_{-0.4}$ & $1.6 \pm 0.1 ^{+0.5}_{-0.4}$ & $5.5 \pm 0.3 ^{+1.6}_{-1.3}$ \\ [\cmsTabSkip]
        \POWHEG (\ttbar) & $1.1 \pm 0.2$ & $1.0 \pm 0.2$ & $3.5 \pm 0.6$ \\
        \POWHEG (\ttbar) + \HERWIGpp & $0.8 \pm 0.2$ & $0.8 \pm 0.2$ & $3.0 \pm 0.5$ \\
        \MGvATNLO (4FS \ttbarbb) & $0.8 \pm 0.2$ & $0.8 \pm 0.2$ & $2.3 \pm 0.7$ \\
        \MGvATNLO (5FS $\ttbar$+jets, FxFx) &  $1.0 \pm 0.1$ & $1.0 \pm 0.1$ & $3.6 \pm 0.3$ \\
    \end{tabular}
\end{table*}

\section{Summary}

The first measurement of the \ttbarbb cross section in the all-jet final state was presented, using $35.9\fbinv$ of data collected in $\Pp\Pp$ collisions at $\sqrt{s}=13\TeV$.
The cross section is first measured in a fiducial region of particle-level phase space by defining two categories of \ttbarbb events, and subsequently this result is corrected to the total phase space.
One of the defined fiducial regions corresponds to ignoring parton-level information, while the other uses parton-level information to identify the particle-level jets that do not originate from the decay of top quarks.
For both definitions, the cross section is measured to be $1.6 \pm 0.1\stat^{+0.5}_{-0.4}\syst\unit{pb}$.
The cross section in the total phase space is obtained by correcting this measurement for the experimental acceptance on the jets originating from the top quarks, which yields $5.5 \pm 0.3\stat^{+1.6}_{-1.3}\syst\unit{pb}$.
This measurement provides valuable input to studies of the $\ttbar\PH$ process, where the Higgs boson decays into a pair of \PQb quarks, and for which the normalization and modelling of the \ttbarbb process represent a leading source of systematic uncertainty.
Furthermore, these results represent a stringent test of perturbative quantum chromodynamics at the LHC.
Predictions from several generators are compared with measurements and found to be smaller than the measured values by a factor of 1.5--2.4, corresponding to 1--2 standard deviations.
This is consistent with previous results for the \ttbarbb cross section and calls for further experimental and theoretical studies of the associated production of top quark pairs and \PQb jets.
\begin{acknowledgments}
We congratulate our colleagues in the CERN accelerator departments for the excellent performance of the LHC and thank the technical and administrative staffs at CERN and at other CMS institutes for their contributions to the success of the CMS effort. In addition, we gratefully acknowledge the computing centers and personnel of the Worldwide LHC Computing Grid for delivering so effectively the computing infrastructure essential to our analyses. Finally, we acknowledge the enduring support for the construction and operation of the LHC and the CMS detector provided by the following funding agencies: BMBWF and FWF (Austria); FNRS and FWO (Belgium); CNPq, CAPES, FAPERJ, FAPERGS, and FAPESP (Brazil); MES (Bulgaria); CERN; CAS, MoST, and NSFC (China); COLCIENCIAS (Colombia); MSES and CSF (Croatia); RPF (Cyprus); SENESCYT (Ecuador); MoER, ERC IUT, PUT and ERDF (Estonia); Academy of Finland, MEC, and HIP (Finland); CEA and CNRS/IN2P3 (France); BMBF, DFG, and HGF (Germany); GSRT (Greece); NKFIA (Hungary); DAE and DST (India); IPM (Iran); SFI (Ireland); INFN (Italy); MSIP and NRF (Republic of Korea); MES (Latvia); LAS (Lithuania); MOE and UM (Malaysia); BUAP, CINVESTAV, CONACYT, LNS, SEP, and UASLP-FAI (Mexico); MOS (Montenegro); MBIE (New Zealand); PAEC (Pakistan); MSHE and NSC (Poland); FCT (Portugal); JINR (Dubna); MON, RosAtom, RAS, RFBR, and NRC KI (Russia); MESTD (Serbia); SEIDI, CPAN, PCTI, and FEDER (Spain); MOSTR (Sri Lanka); Swiss Funding Agencies (Switzerland); MST (Taipei); ThEPCenter, IPST, STAR, and NSTDA (Thailand); TUBITAK and TAEK (Turkey); NASU and SFFR (Ukraine); STFC (United Kingdom); DOE and NSF (USA).

\hyphenation{Rachada-pisek} Individuals have received support from the Marie-Curie program and the European Research Council and Horizon 2020 Grant, contract Nos.\ 675440, 752730, and 765710 (European Union); the Leventis Foundation; the A.P.\ Sloan Foundation; the Alexander von Humboldt Foundation; the Belgian Federal Science Policy Office; the Fonds pour la Formation \`a la Recherche dans l'Industrie et dans l'Agriculture (FRIA-Belgium); the Agentschap voor Innovatie door Wetenschap en Technologie (IWT-Belgium); the F.R.S.-FNRS and FWO (Belgium) under the ``Excellence of Science -- EOS" -- be.h project n.\ 30820817; the Beijing Municipal Science \& Technology Commission, No. Z181100004218003; the Ministry of Education, Youth and Sports (MEYS) of the Czech Republic; the Lend\"ulet (``Momentum") Program and the J\'anos Bolyai Research Scholarship of the Hungarian Academy of Sciences, the New National Excellence Program \'UNKP, the NKFIA research grants 123842, 123959, 124845, 124850, 125105, 128713, 128786, and 129058 (Hungary); the Council of Science and Industrial Research, India; the HOMING PLUS program of the Foundation for Polish Science, cofinanced from European Union, Regional Development Fund, the Mobility Plus program of the Ministry of Science and Higher Education, the National Science Center (Poland), contracts Harmonia 2014/14/M/ST2/00428, Opus 2014/13/B/ST2/02543, 2014/15/B/ST2/03998, and 2015/19/B/ST2/02861, Sonata-bis 2012/07/E/ST2/01406; the National Priorities Research Program by Qatar National Research Fund; the Ministry of Science and Education, grant no. 3.2989.2017 (Russia); the Programa Estatal de Fomento de la Investigaci{\'o}n Cient{\'i}fica y T{\'e}cnica de Excelencia Mar\'{\i}a de Maeztu, grant MDM-2015-0509 and the Programa Severo Ochoa del Principado de Asturias; the Thalis and Aristeia programs cofinanced by EU-ESF and the Greek NSRF; the Rachadapisek Sompot Fund for Postdoctoral Fellowship, Chulalongkorn University and the Chulalongkorn Academic into Its 2nd Century Project Advancement Project (Thailand); the Nvidia Corporation; the Welch Foundation, contract C-1845; and the Weston Havens Foundation (USA).
\end{acknowledgments}

\bibliography{auto_generated}
\cleardoublepage \appendix\section{The CMS Collaboration \label{app:collab}}\begin{sloppypar}\hyphenpenalty=5000\widowpenalty=500\clubpenalty=5000\vskip\cmsinstskip
\textbf{Yerevan Physics Institute, Yerevan, Armenia}\\*[0pt]
A.M.~Sirunyan$^{\textrm{\dag}}$, A.~Tumasyan
\vskip\cmsinstskip
\textbf{Institut f\"{u}r Hochenergiephysik, Wien, Austria}\\*[0pt]
W.~Adam, F.~Ambrogi, T.~Bergauer, J.~Brandstetter, M.~Dragicevic, J.~Er\"{o}, A.~Escalante~Del~Valle, M.~Flechl, R.~Fr\"{u}hwirth\cmsAuthorMark{1}, M.~Jeitler\cmsAuthorMark{1}, N.~Krammer, I.~Kr\"{a}tschmer, D.~Liko, T.~Madlener, I.~Mikulec, N.~Rad, J.~Schieck\cmsAuthorMark{1}, R.~Sch\"{o}fbeck, M.~Spanring, D.~Spitzbart, W.~Waltenberger, C.-E.~Wulz\cmsAuthorMark{1}, M.~Zarucki
% ...
\vskip\cmsinstskip
\dag: Deceased\\
1:  Also at Vienna University of Technology, Vienna, Austria\\
% ...
\end{sloppypar}
\end{document}